\documentclass[runningheads]{llncs}
\usepackage{float}
\usepackage{graphicx} 
\usepackage{amsmath} 
\usepackage{url}
\usepackage{placeins} 
\usepackage{subcaption}
\begin{document}

\title{Combining GPU and CPU for accelerating evolutionary computing workloads}
\author{Rustam Eynaliyev, Houcen Liu}
\date{December 2024}

\authorrunning{Eynaliyev et al.}

\institute{Vrije Universiteit Amsterdam}

\maketitle

\begin{abstract}

Evolutionary computing (EC) has proven to be effective in solving complex optimization and robotics problems. Unfortunately, typical Evolutionary Algorithms (EAs) are constrained by the computational capacity available to researchers. More recently, GPUs have been extensively used in speeding up workloads across a variety of fields in AI. This led us to the idea of considering utilizing GPUs for optimizing ECs, particularly for complex problems such as the evolution of artificial creatures in physics simulations.
In this study, we compared the CPU and GPU performance across various simulation models, from simple box environments to more complex models. Additionally, we create and investigate a novel hybrid CPU + GPU scheme that aims to fully utilize the idle hardware capabilities present on most consumer devices. The strategy involves running simulation workloads on both the GPU and the CPU, dynamically adjusting the distribution of workload between the CPU and the GPU based on benchmark results. Our findings suggest that while the CPU demonstrates superior performance under most conditions, the hybrid CPU + GPU strategy shows promise at higher workloads. However, overall performance improvement is highly sensitive to simulation parameters such as the number of variants, the complexity of the model, and the duration of the simulation. These results demonstrate the potential of creative, dynamic resource management for experiments running physics simulations on workstations and consumer devices that have both GPUs and CPUs present.
\end{abstract}

\section{Introduction} 
With AI increasingly entering our lives in crucial social environments 
(e.g., healthcare, education), efforts to speed up its development become 
ever more important.~\cite{medgyesy2025}.
Evolutionary computing (EC) applies the idea of natural selection to solve challenging problems in robotics, optimization, and constraint satisfaction in large /complex problem spaces \cite{eiben2015}(pp. 10--12). Evolutionary algorithms have introduced significant contributions to real-world problems including scheduling and redesigning of factory production flow, design of antennas \cite{hornby2011}, computer programs, \cite{koza2010}  artificial life \cite{eiben2015nature}, robot morphologies, \cite{eiben2015} (pp. 22--23).

However, such goals are not without a great amount of work, as the fitness function can be noisy or expensive, the environment may be difficult to navigate, and goals can be constantly changing while the algorithm must remain robust \cite{eiben2015} (Chapter 17, pp. 251–252). Researchers have responded by exploring strategies such as novelty-driven exploration for uncharted environments \cite{eiben2015} (Chapter 17, p. 252) and combining multiple objectives to evolve more complex multimodal behaviors \cite{huizinga2012}.

At the core of many advances in the field of EC remains the need to reduce computational requirements. The use of high-fidelity physics-based models for robot simulation is highly computationally demanding. Moreover, efforts to decrease the "reality gap", or the discrepancy between simulation and physical reality, require us to use more accurate and, therefore, more computationally complex simulators \cite{eiben2015}(Chapter 17, p. 253).

These problems become ever more prevalent in memetic algorithms where every generation in the Evolution Algorithm has a local search further applied to it.\cite{eiben2015} This typically results in a significant computational increase as every candidate solution further undergoes multiple variance and evaluation steps. Memetic algorithms are of particular interest in the quest of designing artificial creatures where both the Robot morphology and controller can be evolved independently, enabling for adaptation of the Robot's controller to its morphology. A variant of this approach, Lamarckian evolution, has been successfully demonstrated to significantly enhance the performance of EAs. \cite{luo2023}

Several works underscore the need for the utilization of GPUs to address these computational challenges. For instance, frameworks designed for distributed GPU acceleration, such as EvoX (\cite{huang2023}) directly solve the scalability issue since the large-scale evolutionary computation is carried out in parallel. This shows that GPU–based solutions can make a difference in large-scale evolutionary runs in terms of both speed and adaptability.

In this paper, our aim is to optimize the framework for Evolutionary algorithms, Revolve2 \cite{revolve2}. Revolve2 is used for the design of artificial creatures and is based on the Mujoco simulator \cite{todorov2012}. In order to do that, we aim to benchmark Mujoco against its GPU-optimized variant, Mujoco XLA (MJX) \cite{mjx2023}.
By improving the runtime of EAs we aim to further accelerate the pace of scientific progress by enabling faster experimentation for researchers.

This paper is organized as follows. In Section 1, the reader is provided with an understanding of why GPU-based MJX simulations are relevant in Evolutionary Computing and the research questions that guide the study. Section 2 describes the experimental procedure and all the tools and resources employed in this work. Section 3 then describes the initial profiling of an example run of the evolutionary algorithm. Section 4 presents our benchmarking of the CPU against the GPU for different numbers of simulation variants and models. In Section 5, we analyze the dependence of GPU runtime on the number of simulation steps and explain how resource saturation affects expected speedups. Section 6 presents an example of the dynamic CPU-GPU workload distribution and shows whether adaptive parallel processing can compensate for overhead costs. Section 7 discusses future work, which includes the generalization of parallelism to other higher levels of hardware and a larger scale of simulation. Section 8 concludes by summarizing the main results and stressing the potential of hybrid strategies for enhancing EC-based robotic simulations. 

\section{Methodology}

\subsubsection{Code}

All of the code used for the experiments as well as the benchmarking scripts,models, profiling and visualization scripts as well as experiment results described in this paper are publicly available on GitHub at:

\url{https://github.com/rustam-e/revolve2/tree/mujoco-profile-enhanced/mjx_profile_experiments}

\subsection{Materials}
\subsubsection{Hardware and Software}

Experiments were conducted using the following hardware and software:

\begin{itemize}
    \item \textbf{Hardware}:
    \begin{itemize}
        \item AMD Ryzen Threadripper 2990WX 32-Core Processor
        \item GPU: NVIDIA GeForce GTX 1070 Ti with 10GB GDDR6X VRAM
        \item RAM: 64GB
    \end{itemize}
    \item \textbf{Software}:
    \begin{itemize}
        \item Operating System: Ubuntu 22.04.5 LTS
        \item Programming Language: Python 3.10
        \item Physics Engine: MuJoCo 3.2.6 for CPU simulations and MJX for GPU simulations support~\cite{todorov2012}
        \item Libraries: NumPy, Matplotlib, psutil, NVIDIA CUDA Toolkit, SnakeViz
    \end{itemize}
\end{itemize}

\subsection{Profiling Tools Used}

\begin{enumerate}
    \item \textbf{Python's \texttt{cProfile}}:
    \begin{itemize}
        \item Profiler used for detecting performance bottlenecks in example EC algorithms. Was used in initial exploratory analysis. 
        \item Visualized using \texttt{SnakeViz}, a web-based tool for visual analysis.
    \end{itemize}
    \item \textbf{NVIDIA \texttt{nvidia-smi}}:
    \begin{itemize}
        \item Command-line utility for monitoring GPU utilization, memory usage, and performance metrics.
    \end{itemize}
    \item \textbf{Custom Benchmarking Scripts}:
    \begin{itemize}
        \item Allowed comparing hybrid combined strategy to using CPU or GPU standalone. 
    \end{itemize}
\end{enumerate}

\section{Initial profiling of Evolutionary Algorithm run}

We started by profiling the runtime of example evolutionary algorithms from the Revolve2 example library running on the CPU. The investigation showed that over 80 percent of the runtime of the algorithm was spent during the actual physics simulation part of the algorithm.

For profiling, we use Python’s cProfile module. For visualizing the output of Python’s cProfile module we use SnakeViz.

\begin{figure}[htbp]
    \centering
    \includegraphics[width=\linewidth]{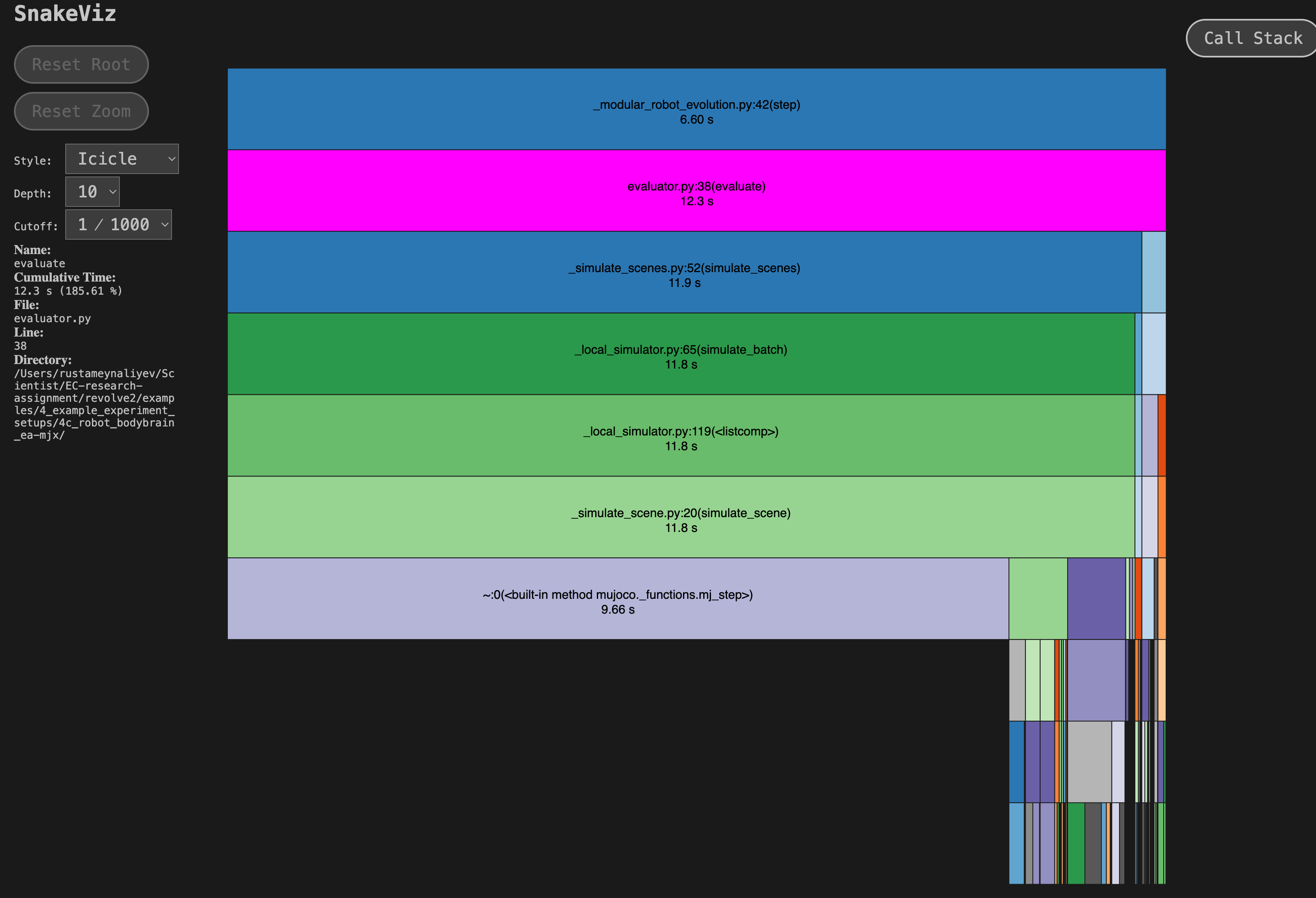} 
    \caption{Profile statistics of runtime of an Evolutionary Algorithm}
    \label{fig:profile_stats}
\end{figure}

\FloatBarrier 

\section{Initial Benchmarking: CPU vs GPU performance}

Based on the results of the initial profile, our follow-up efforts have focused on updating the underlying Mujoco simulator to MJX \cite{mjx2023} simulator, which is specifically built to add GPU support to Mujoco.

In order to do that, we build a simple Mujoco simulation benchmarking script that tests a set of various models, simulation duration, and variant combinations.

\subsubsection{Simulation Parameters}

The experiments were conducted using the following parameters:

\begin{itemize} 
    \item \textbf{Simulation variants}: 
        \begin{itemize} 
        \item \textit{BOX}: 32, 128, 256, 512, 1024, 2056, 4096, 8192, 16384, 32768, 65536, 131072, 256000, 512000 
        \item \textit{BOX\_AND\_BALL}: 32, 128, 256, 512, 1024, 2056, 4096, 8192, 16384, 32768, 65536, 131072, 256000, 512000 
        \item \textit{ARM\_WITH\_ROPE}: 32, 128, 256, 512, 1024, 2056, 4096, 8192, 16384, 32768, 65536, 131072, 256000 
        \item \textit{HUMANOID}: 32, 128, 256, 512, 1024, 2056, 4096, 8192, 16384, 32768 
        \end{itemize} 
    \item \textbf{Simulation Steps}: 1000 
    \item \textbf{Repetitions}: 3 
\end{itemize}

The differences in the number of variants per model are due to memory limitations encountered during the simulations. For models requiring more memory, the maximum number of variants was reduced to ensure successful execution of the experiments.

\subsection{Code}
We created a benchmarking script to automate performance tests across different configurations. We ran each experiment on CPU using Mujoco and on GPU using MJX.
We used \texttt{psutil} and NVIDIA's \texttt{nvidia-smi} for monitoring resource utilization. 
Additionally, a variant of the combined adaptive GPU / CPU allocation was created. The combined variant runs the simulation once sequentially and then on the second run allocates variants across CPU/GPU proportionately to their respective performance.

All of the code can be found at \url{https://github.com/rustam-e/revolve2/tree/mujoco-profile-enhanced/mjx_profile_experiments}

\subsection{Results}
\label{sec:results}

\begin{figure}[htbp]
    \centering
    \begin{minipage}{0.45\textwidth}
        \centering
        \includegraphics[width=\textwidth]{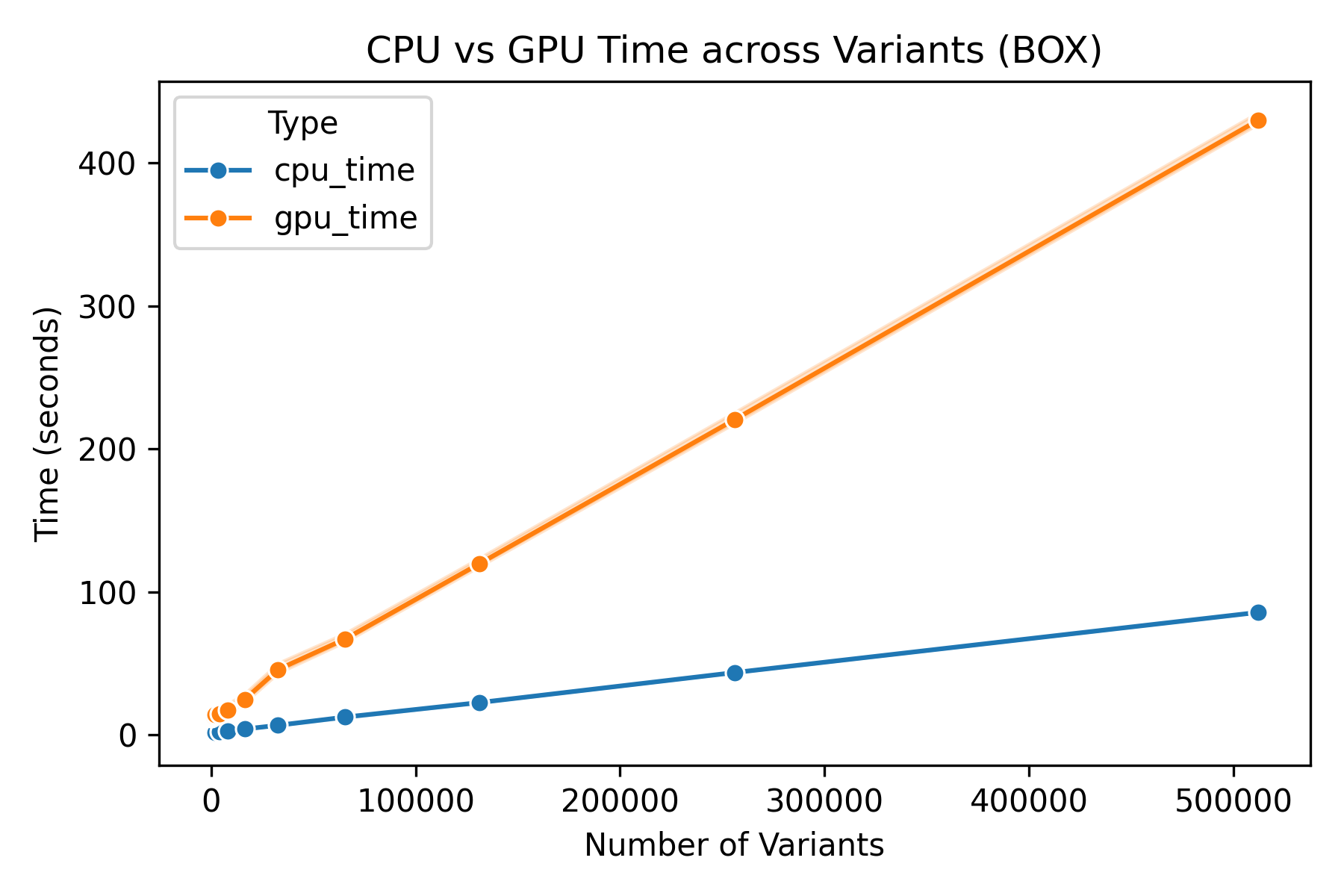}
        \label{fig:cpu_gpu_box}
    \end{minipage}
    \hfill
    \begin{minipage}{0.45\textwidth}
        \centering
        \includegraphics[width=\textwidth]{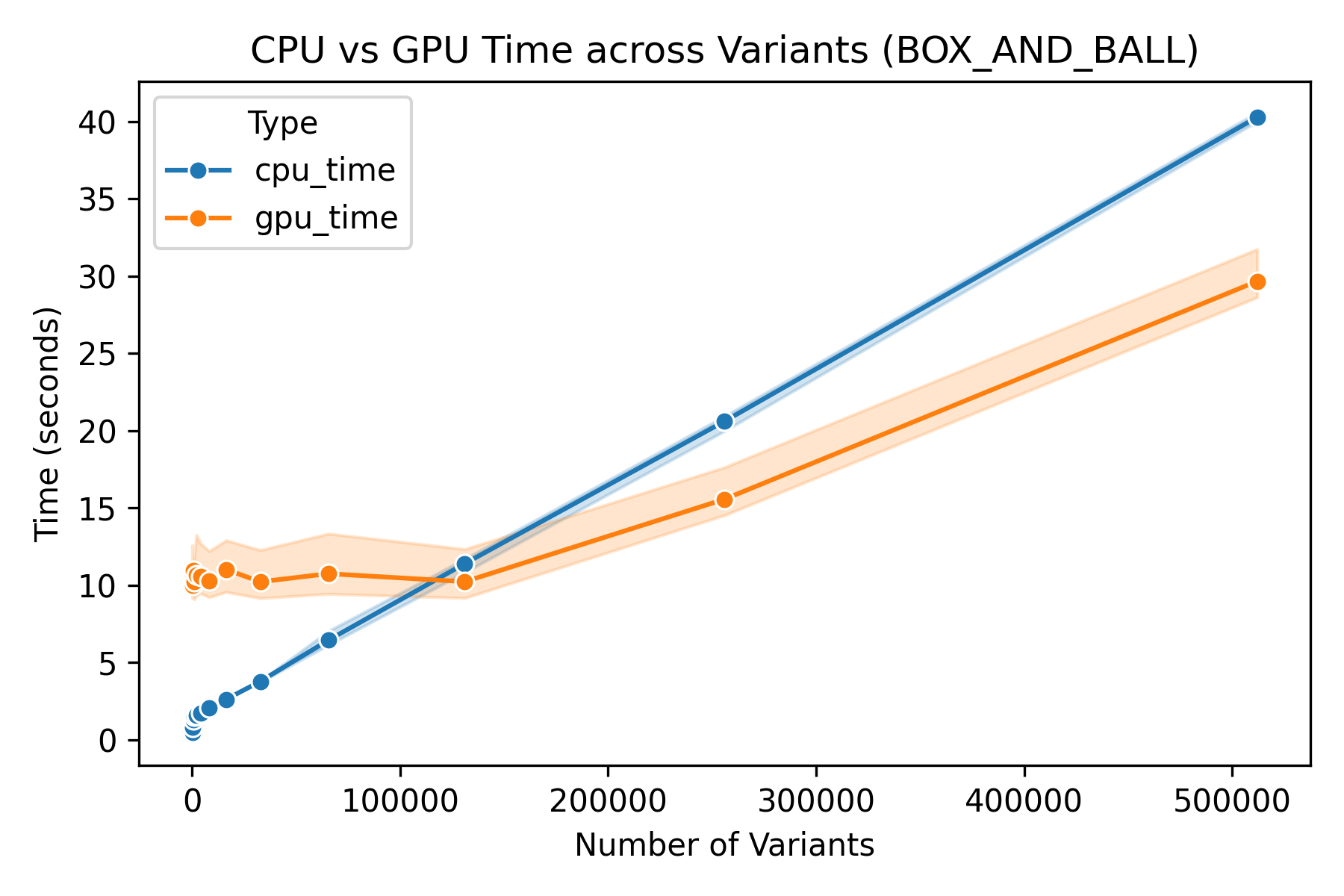}
        \label{fig:cpu_gpu_box_and_ball}
    \end{minipage}
    \vspace{1em}
    \begin{minipage}{0.45\textwidth}
        \centering
        \includegraphics[width=\textwidth]{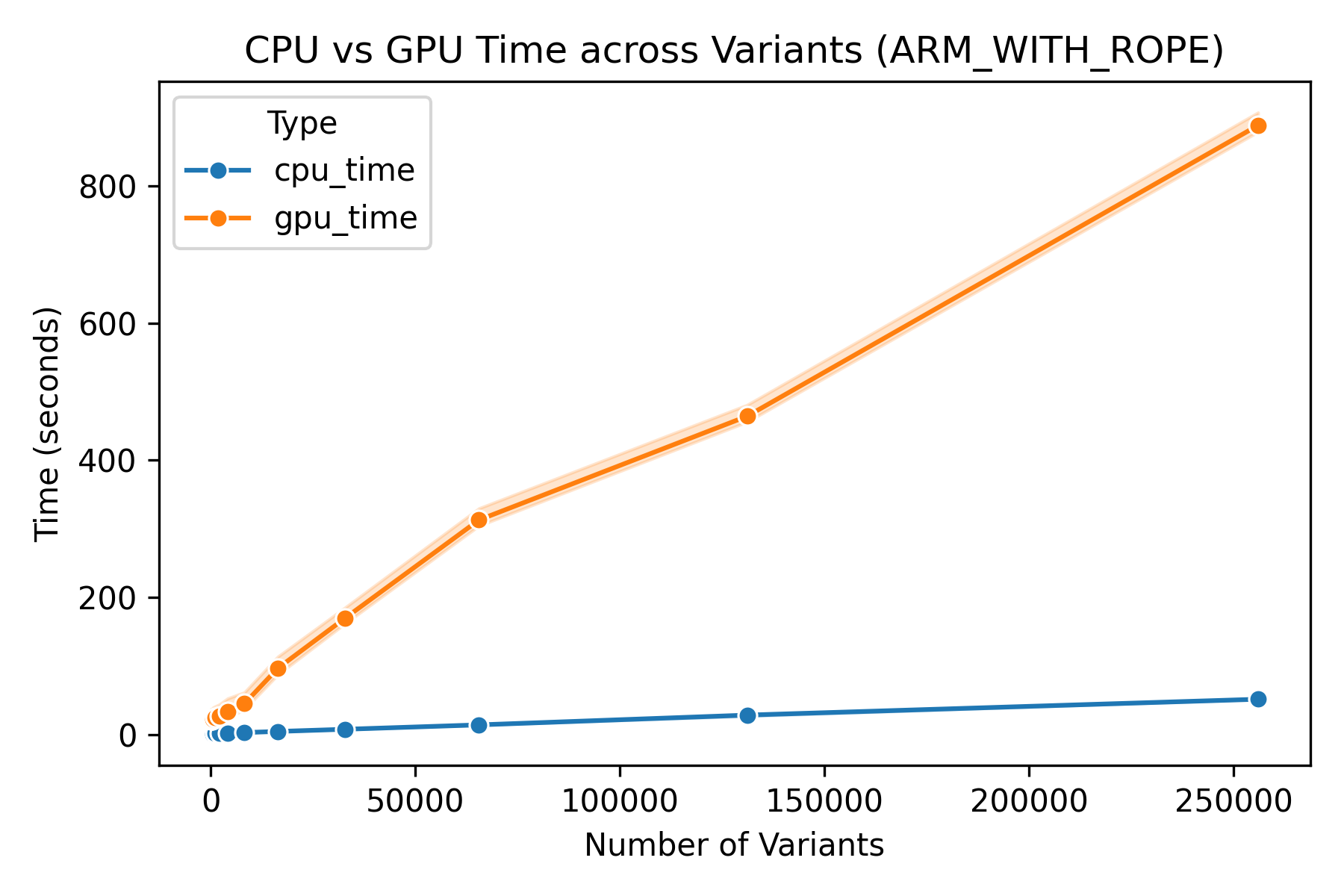}
        \label{fig:cpu_gpu_arm_with_rope}
    \end{minipage}
    \hfill
    \begin{minipage}{0.45\textwidth}
        \centering
        \includegraphics[width=\textwidth]{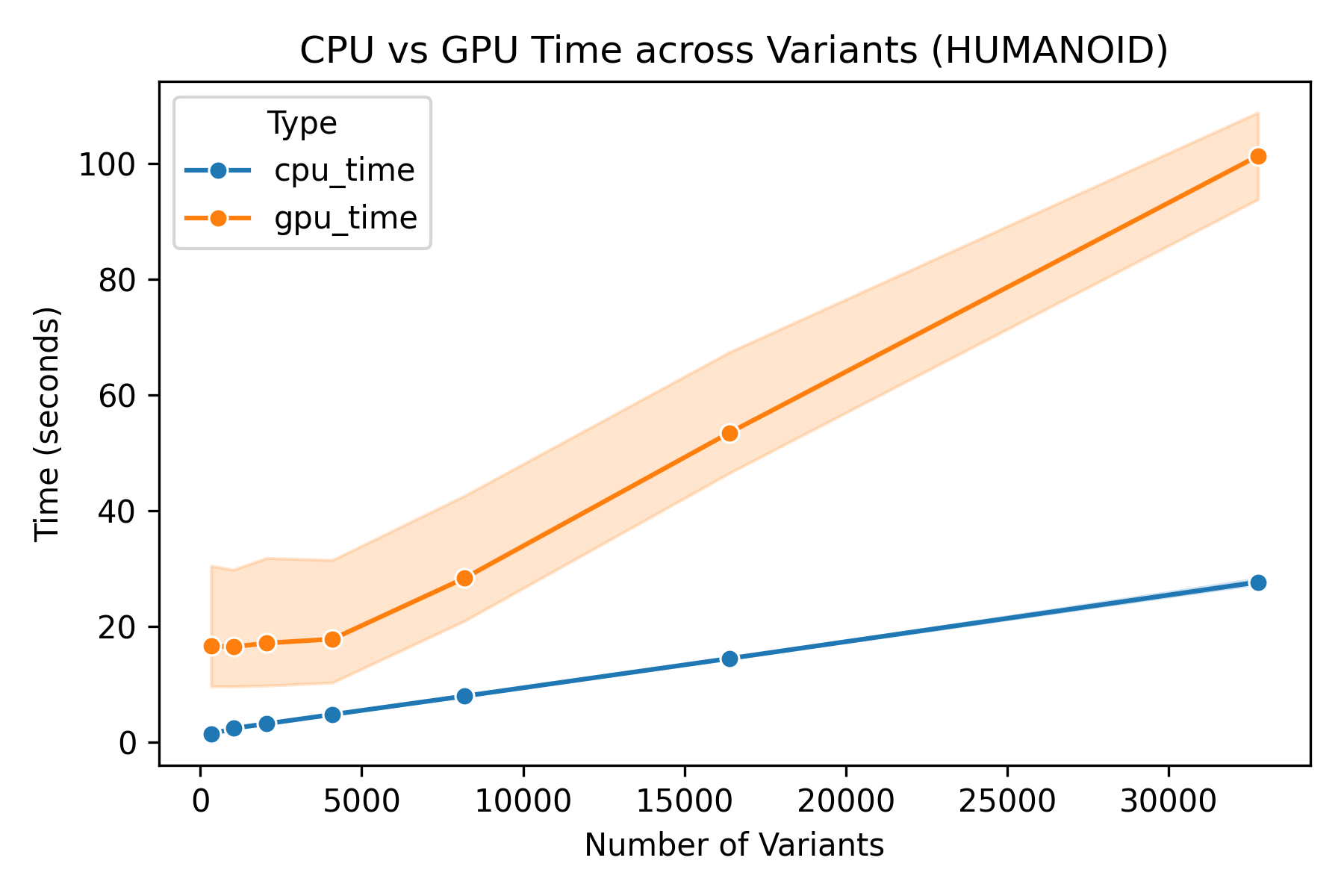}
        \label{fig:cpu_gpu_humanoid}
    \end{minipage}
    \caption{CPU vs. GPU execution time for different simulation models across a range of variants. Shaded regions represent the 95\% confidence interval after 3 repetitions.}
    \label{fig:cpu_gpu_variants}
\end{figure}

This section presents the results of our experiments, comparing the performance of GPU and CPU simulations across a range of evolutionary computing scenarios. The results focus on key metrics such as execution time, speed differences, and resource utilization for different models and numbers of simulation variants. 
Figure~\ref{fig:cpu_gpu_variants} compares the execution time of GPU and CPU simulations for different models and numbers of variants. The graphs highlight the following trends:

\begin{itemize}
    \item \textbf{General Performance Trends:}:Across wide range of numbers of variants, the CPU often outperforms the GPU  with only exception being noted in the \textit{BOX\_AND\_BALL} simulation after around 120,000 variants. 
    \item \textbf{GPU performance variance}: Across all models (e.g., \textit{BOX}, \textit{BOX\_AND\_BALL}, \textit{ARM\_WITH\_ROPE}, and \textit{HUMANOID}), GPU execution time shows significant variability
    \item \textbf{Difference between simulations}:\textit{HUMANOID}) simulation had much higher variance in GPU runtimes than other models
\end{itemize}

\FloatBarrier 
\subsection{GPU Utilization Analysis}

\begin{figure}[htbp]
    \centering
    \begin{minipage}{0.45\textwidth}
        \centering
        \includegraphics[width=\textwidth]{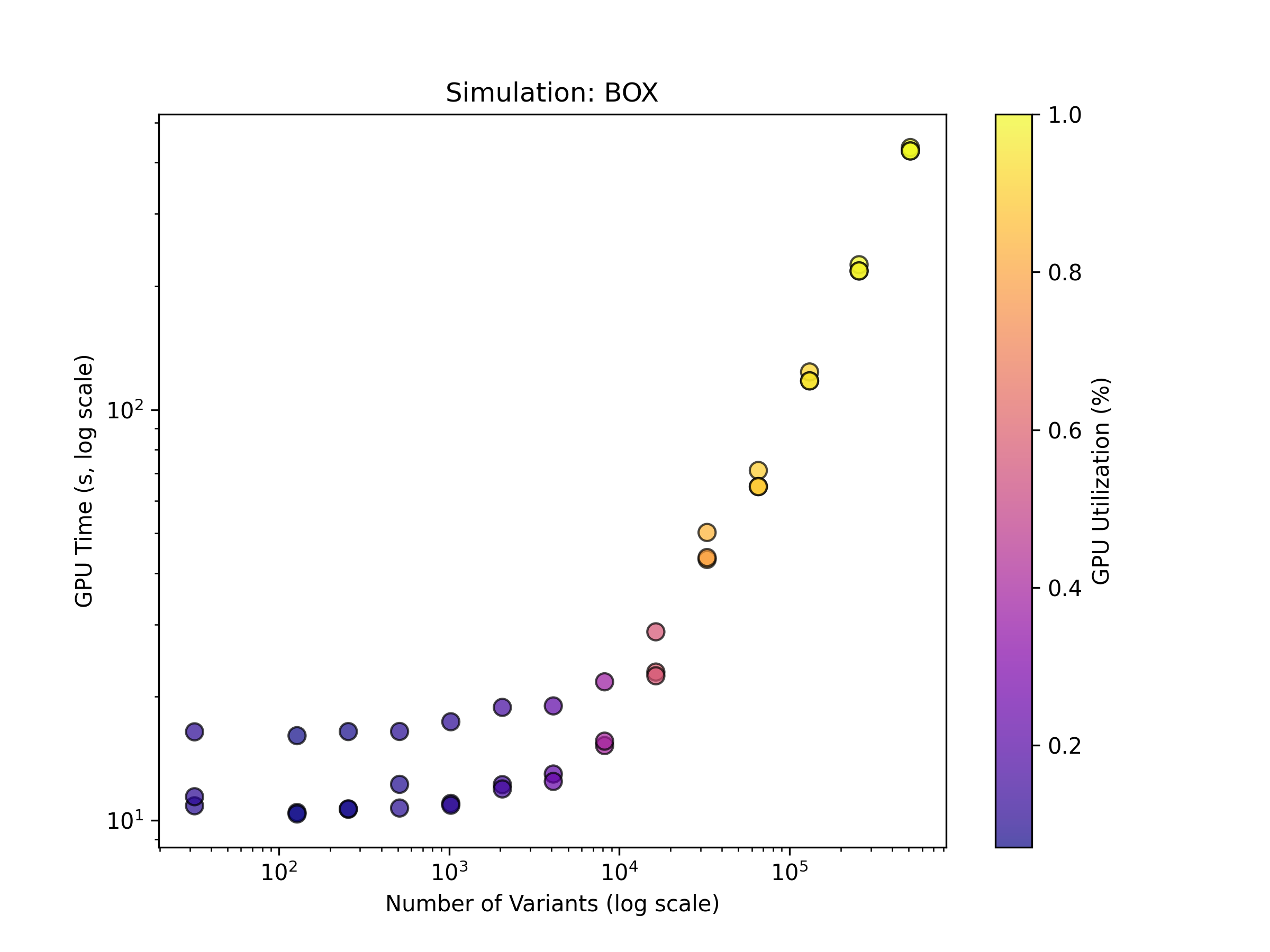}
    \end{minipage}
    \hfill
    \begin{minipage}{0.45\textwidth}
        \centering
        \includegraphics[width=\textwidth]{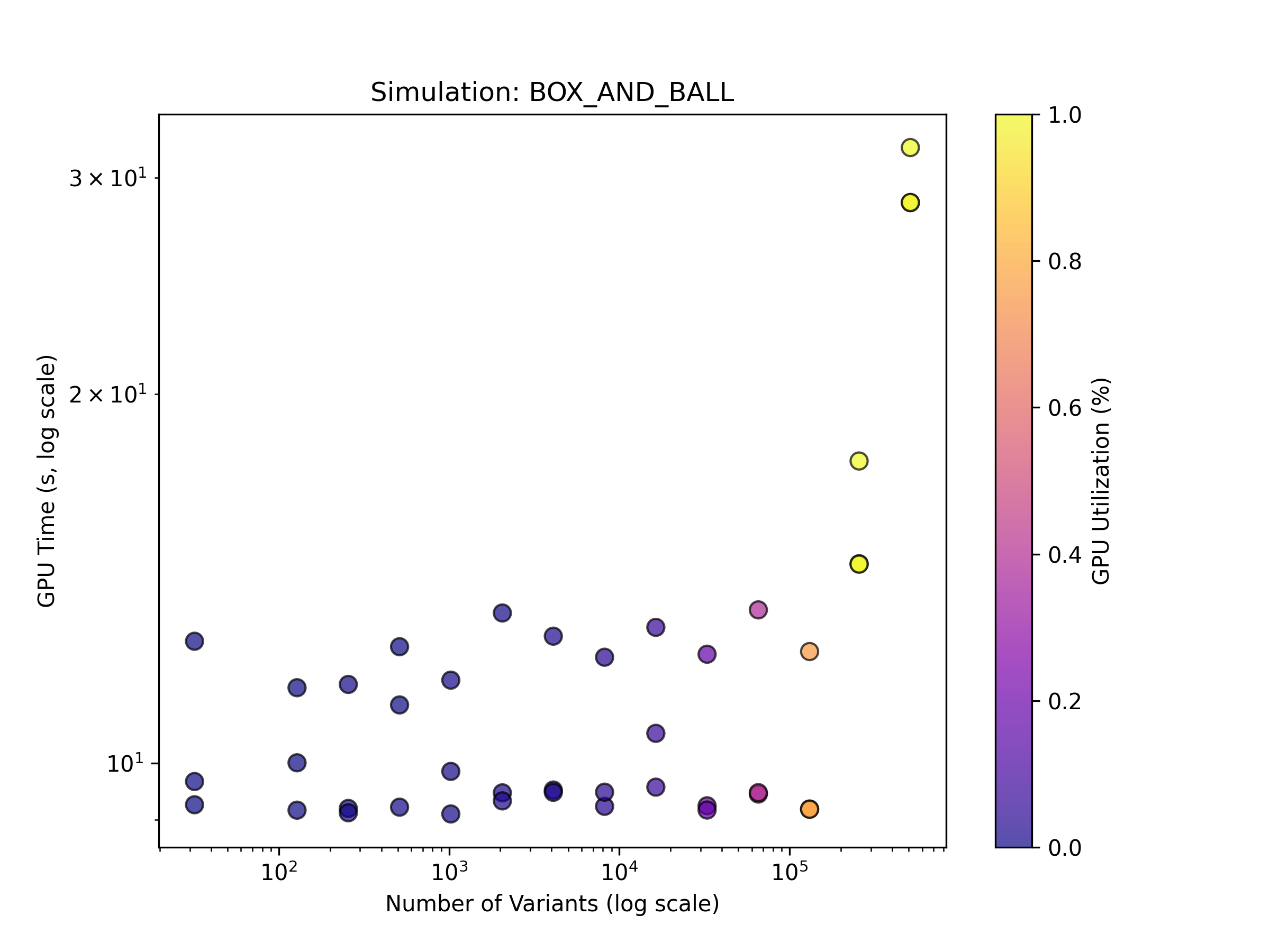}
    \end{minipage}
    \vspace{1em}
    \begin{minipage}{0.45\textwidth}
        \centering
        \includegraphics[width=\textwidth]{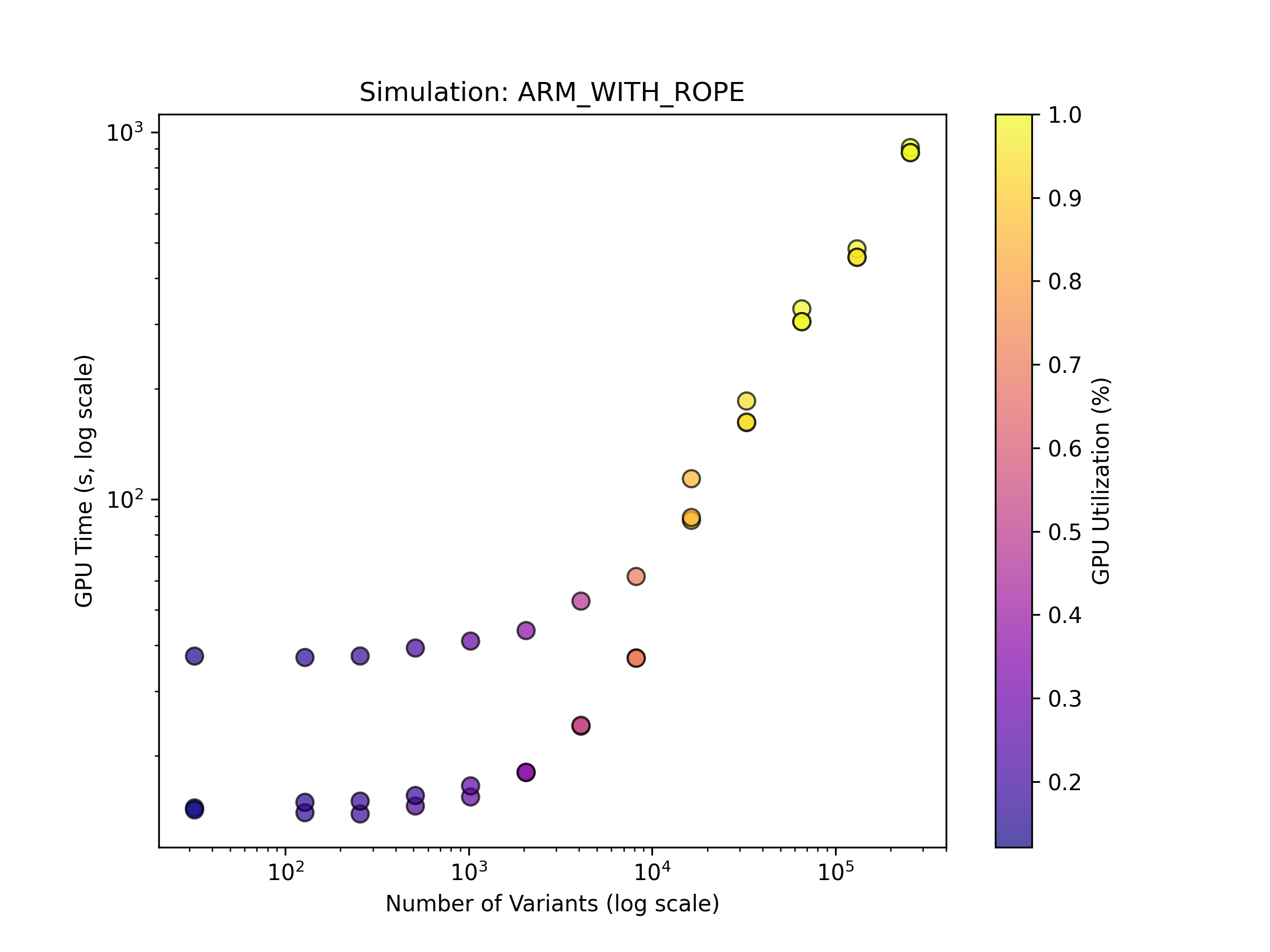}
    \end{minipage}
    \hfill
    \begin{minipage}{0.45\textwidth}
        \centering
        \includegraphics[width=\textwidth]{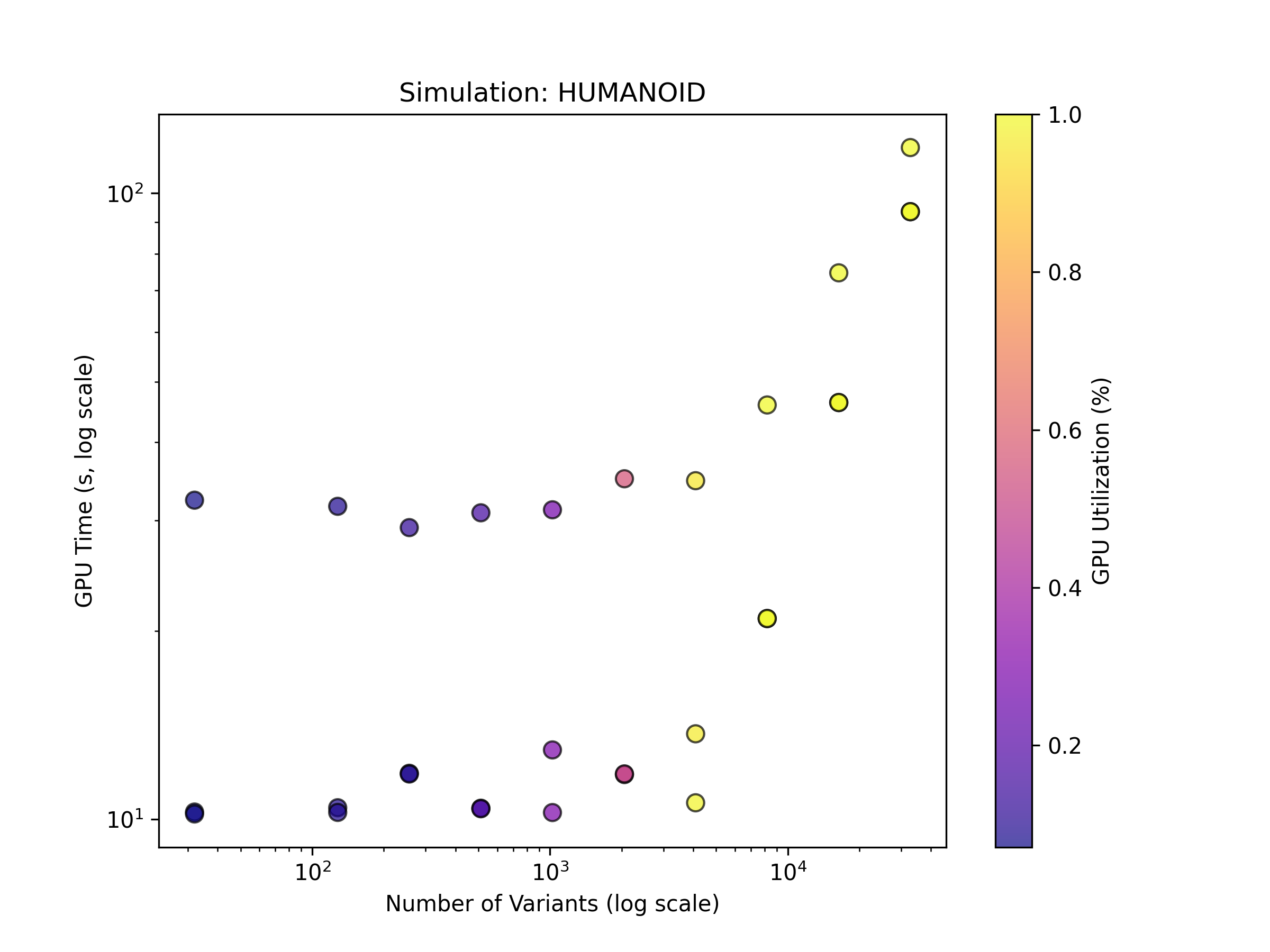}
    \end{minipage}
    \caption{GPU utilization trends across different simulation models as a function of the number of variants. Color indicates GPU utilization (\%).}
    \label{fig:gpu_utilization}
\end{figure}

We next benchmarked the impact of GPU utilization on GPU time. 
Figure~\ref{fig:gpu_utilization} illustrates the GPU time and utilization as a function of the number of variants between models. 

\begin{itemize}
    \item \textbf{Runtime scaling as a function of GPU utilization}: While the GPU utilization is under 100 percent, the runtime of the algorithm stays constant but quickly turns linear once the gpu utilization achieves full capacity.
\end{itemize}

\FloatBarrier 
\section{Benchmarking GPU Runtime Across Simulation Steps}

We continued benchmarking the GPU performance with different number of simulation steps durations. Number of simulation steps is a measure of the duration of individual simulation.
\subsection{Simulation Parameters} 

\begin{itemize}
    \item \textbf{Simulation variants}: \textit{BOX}, \textit{BOX\_AND\_BALL}, \textit{ARM\_WITH\_ROPE}, and \textit{HUMANOID}
    \item \textbf{Population Sizes (Variants)}: 128
    \item \textbf{Simulation Steps}: 32, 128, 256, 512, 1024, 2056, 4096, 8192, 16384, 32768
    \item \textbf{Repetitions}: 3 
\end{itemize}

\subsection{Results}

Key observations:
\begin{itemize}
    \item \textbf{Runtime scaling as a function of GPU utilization}: Just as in the previous test of different variants, here we test the performance of GPU with different simulation lengths. The same as before, while the GPU utilization is under 100 percent, the runtime of the algorithm stays constant, but quickly turns linear once the GPU utilization achieves full capacity.
    \item \textbf{GPU underperformance}: scaling showed similar pattern of GPU underperformance across simulations. 
    \item \textbf{CPU performance}: CPU resulted in significantly performance on longer duration simulations.
    \item \textbf{Increased GPU p95}: The p95 variance after 3 runs increased with the increased number of variants
    \item \textbf{Difference between simulations}: \textit{ARM\_WITH\_ROPE} simulation had much higher variance in GPU runtimes than other models
\end{itemize}

\begin{figure}[htbp]
    \centering
    \begin{minipage}{0.45\textwidth}
        \centering
        \includegraphics[width=\textwidth]{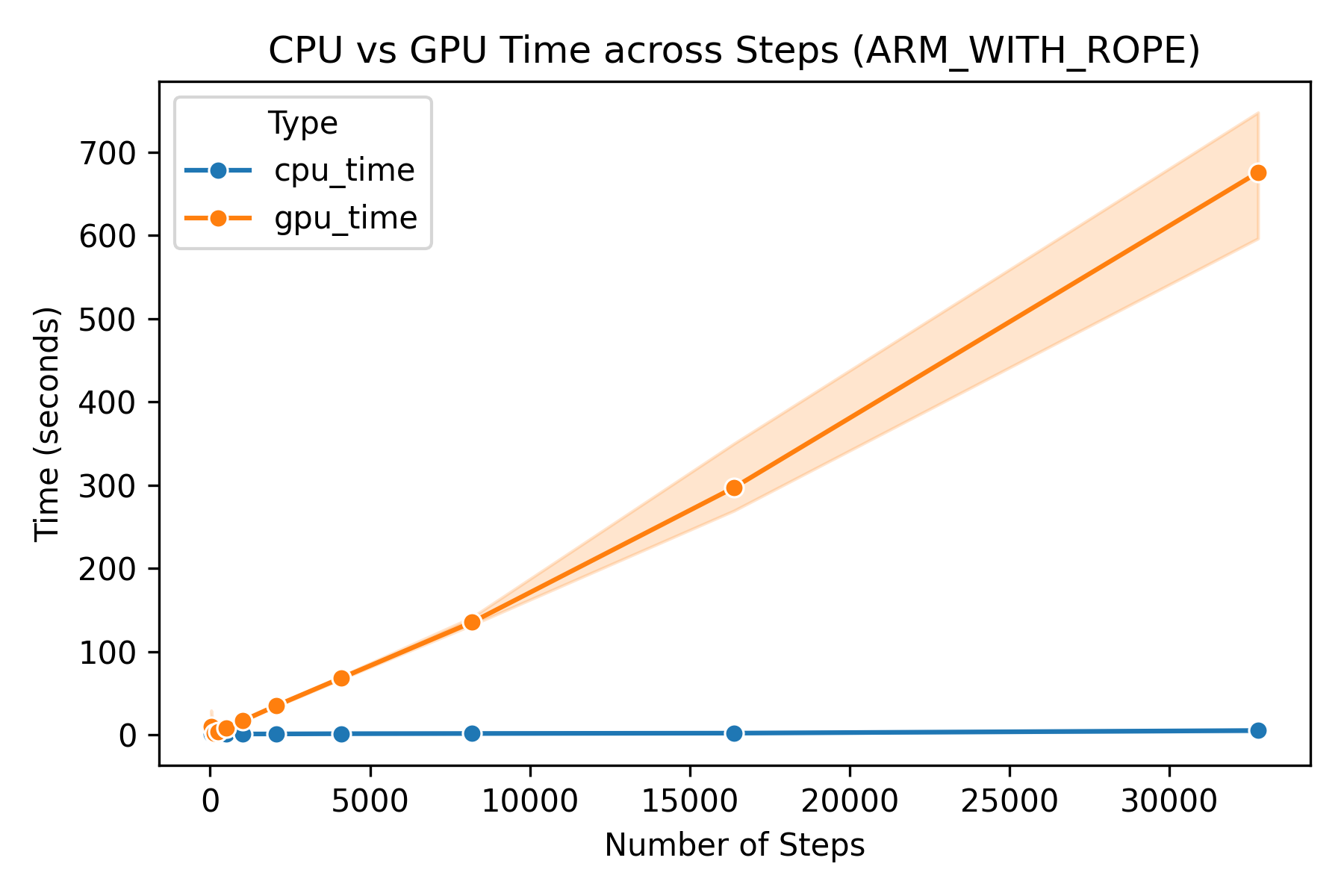}
        \label{fig:multi_step_arm_with_rope}
    \end{minipage}
    \hfill
    \begin{minipage}{0.45\textwidth}
        \centering
        \includegraphics[width=\textwidth]{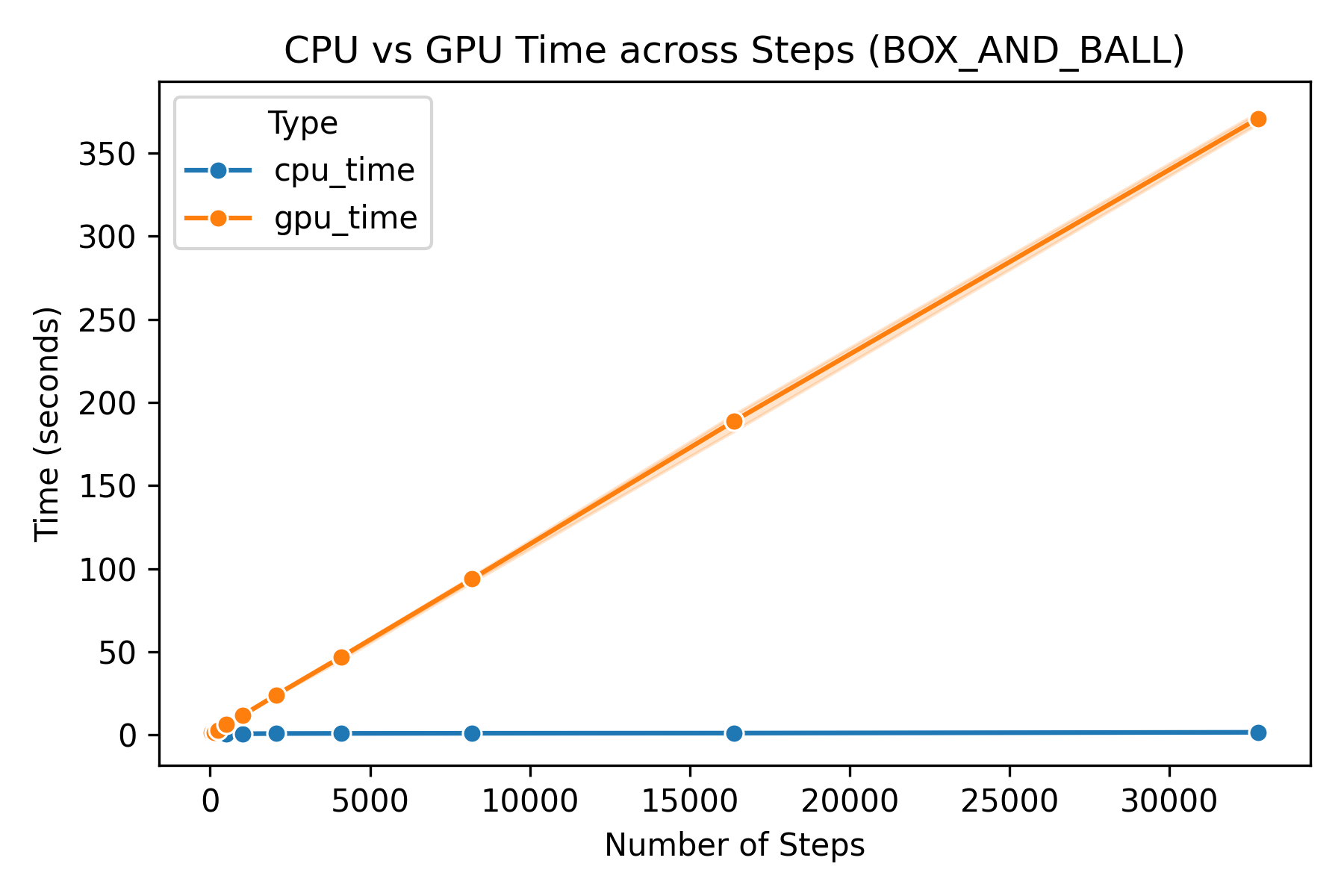}
        \label{fig:multi_step_box_and_ball}
    \end{minipage}
    \vspace{1em}
    \begin{minipage}{0.45\textwidth}
        \centering
        \includegraphics[width=\textwidth]{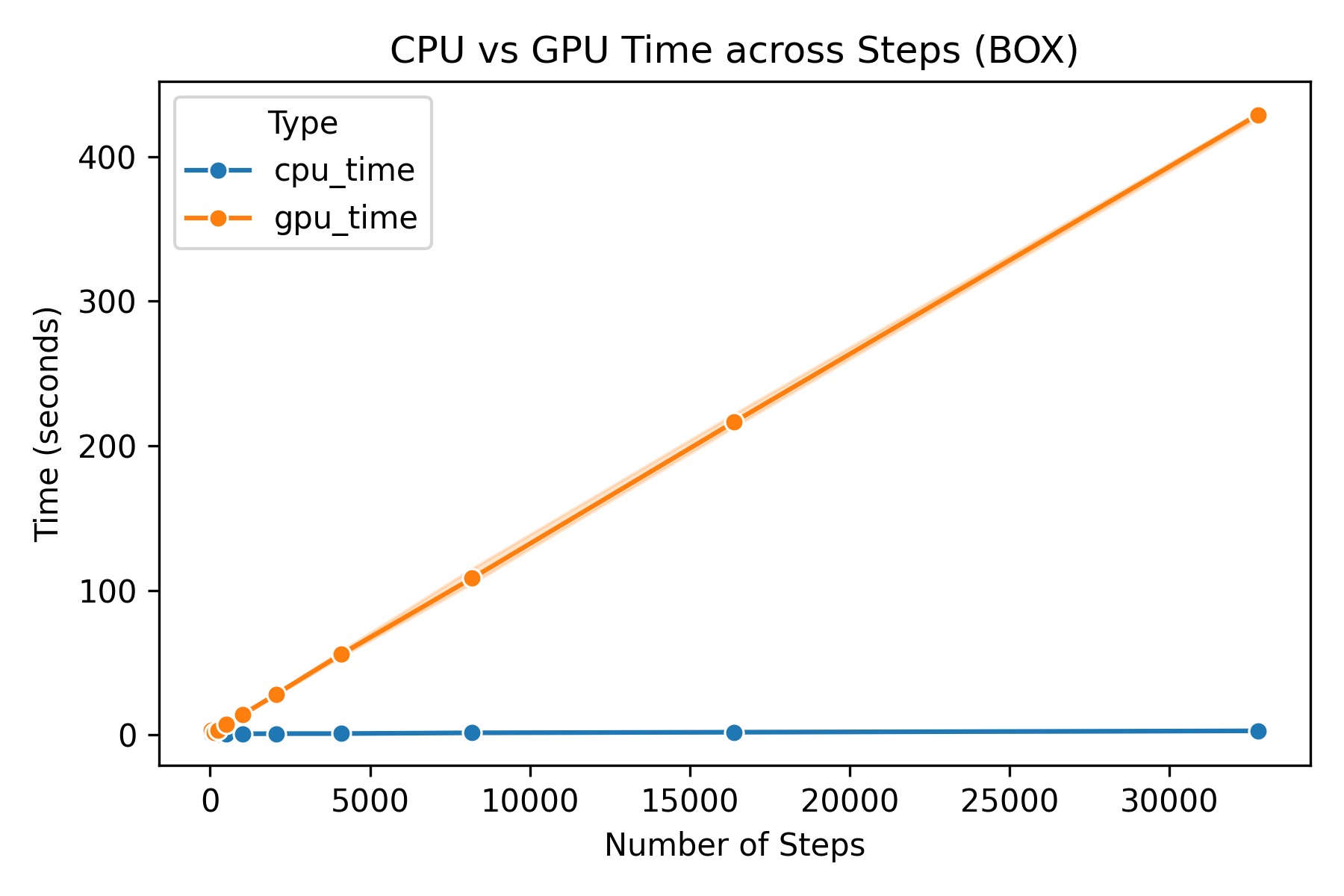}
        \label{fig:multi_step_box}
    \end{minipage}
    \hfill
    \begin{minipage}{0.45\textwidth}
        \centering
        \includegraphics[width=\textwidth]{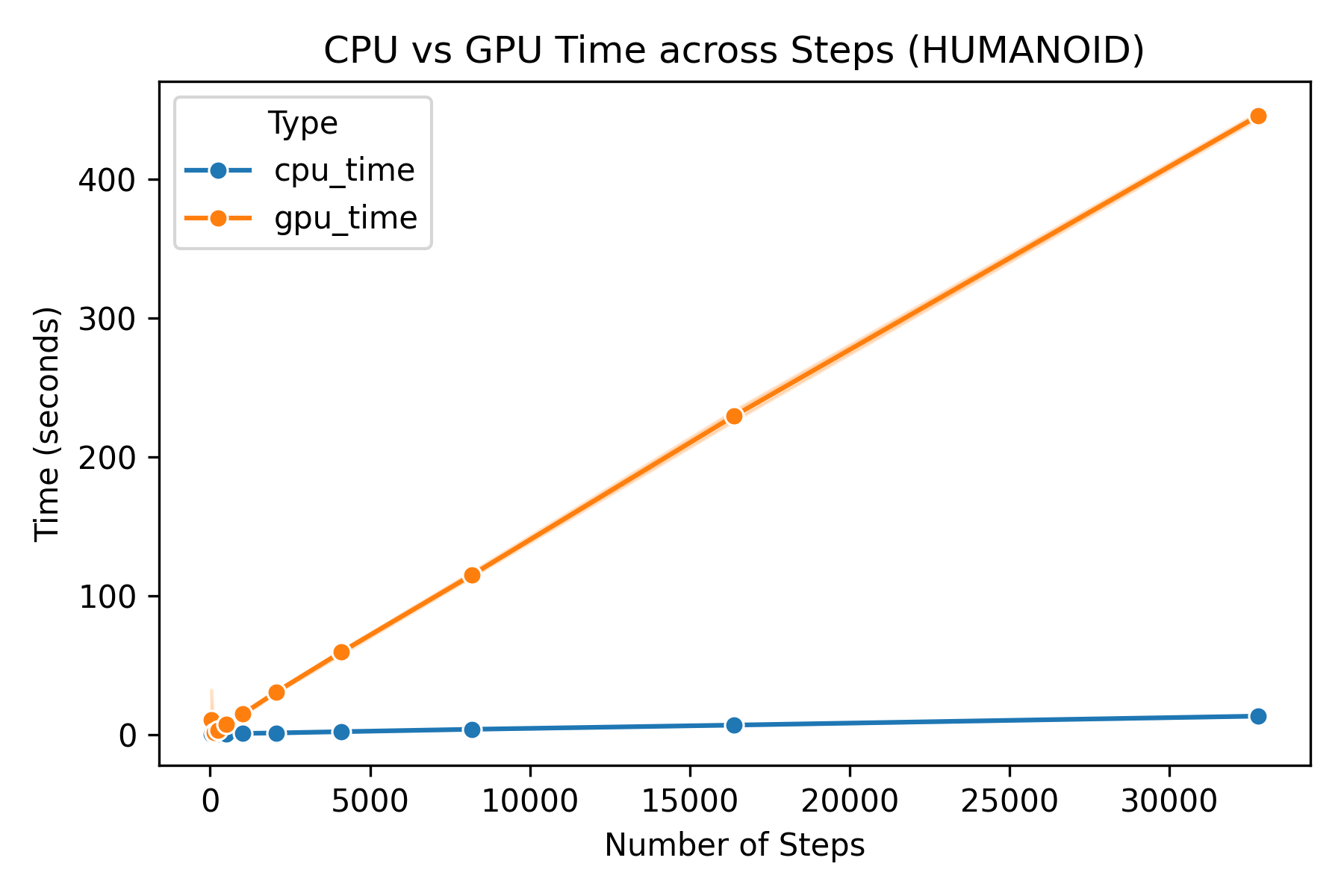}
        \label{fig:multi_step_humanoid}
    \end{minipage}
    \caption{GPU Time vs. Number of Steps for 124 variants (32, 128, 256, 512, 1024, 2056, 4096, 8192, 16384, 32768)}.
    \label{fig:multi_step_gpu_variants}
\end{figure}
\FloatBarrier 
\section{Dynamic Allocation of Variants Across GPU and CPU}
\subsection{Overview}

\begin{figure}[htbp]
\centering
\includegraphics[width=0.8\textwidth]{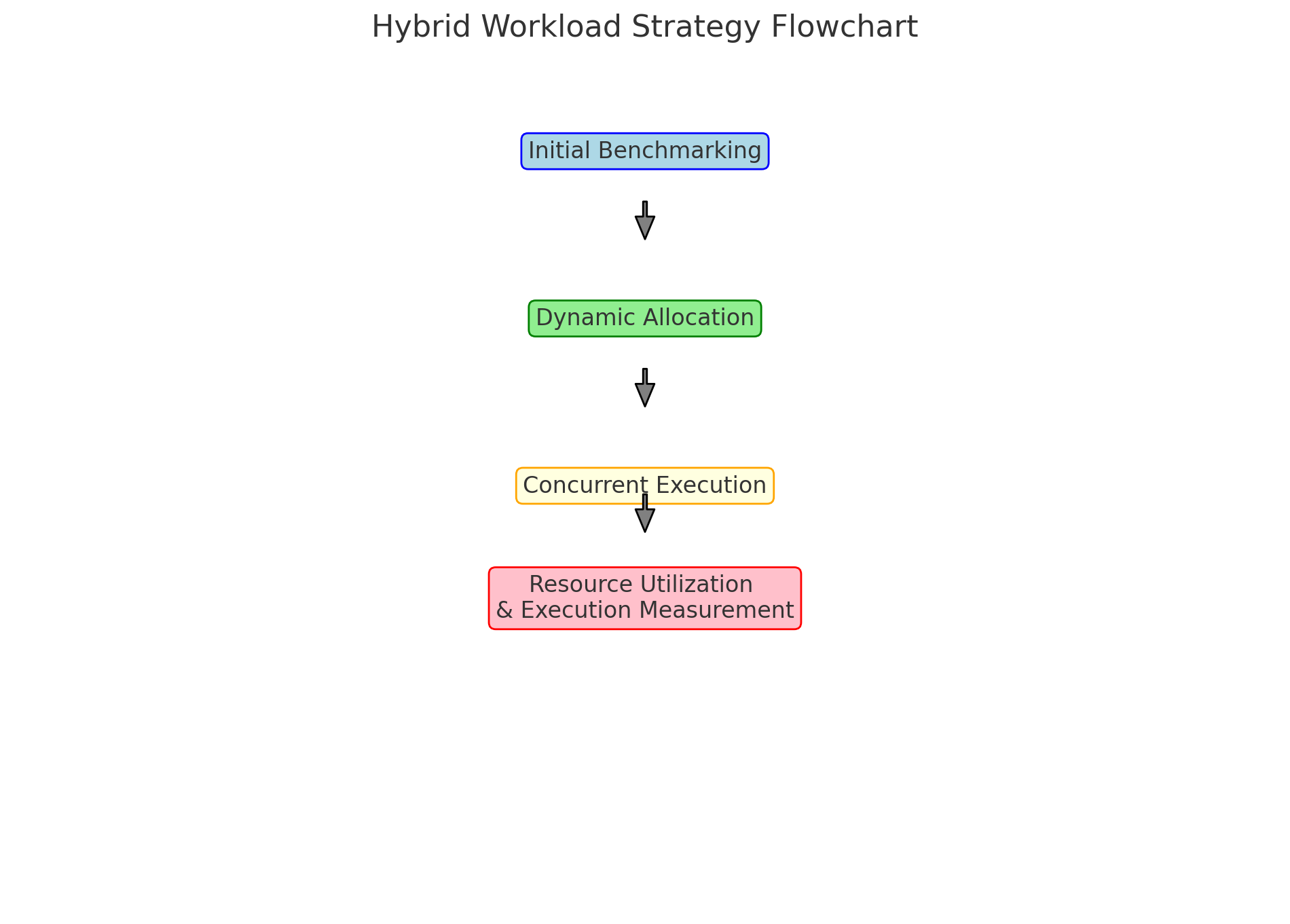}
\caption{Flowchart of the hybrid CPU-GPU workload distribution strategy. Tasks are allocated based on performance metrics gathered during initial benchmarking of running CPU and GPU individually and comparing the ratio of the run times.}
\label{fig:hybrid_strategy}
\end{figure}

After our initial results showed that the CPU vastly outperforms GPUs under almost all conditions, we have implemented a hybrid strategy to offload some of the CPU workload onto the GPU in order to achieve faster wall clock performance. The motivation for this approach was the access to a single desktop machine with both a CPU and a GPU - where using only 1 of the 2 would render the second device idle. We aim to leverage this spare capacity.
\\\\
We hypothesized that the combined allocation of workload across CPU and GPU runtime would result in simulations that are faster than either the CPU or GPU running standalone, and that wall clock at high variants would be faster than the naive summation of CPU and GPU run times.
\\\\
We have designed the following experiment:
A variant of combined adaptive GPU / CPU allocation was created. The combined variant runs the simulation once sequentially and then on the second run allocates variants across CPU/GPU proportionately to their respective performance.
Figure~\ref{fig:hybrid_strategy} illustrates the approach.
\\\\
The strategy can be broken into the following steps:

\begin{enumerate}
    \item \textbf{Initial Benchmarking}: Running combinations of variants, scenes, and number of steps sequentially, tracking the performance of CPU and GPU for each combination. Recording the ratio of GPU time to CPU time.
    \item \textbf{Dynamic Allocation}: Allocating variants across CPU and GPU based on benchmarking results, aiming to minimize total execution time. Reverse the ratio in order to allocate variants across CPU and GPU.
    \item \textbf{Concurrent Execution}: Running simulations concurrently on CPU and GPU. Recording results of actual wallclock runtime of the execution as well as measuring individual run times and resource utilization for CPU and GPU.
    \item \textbf{Resource Utilization Execution}: Measuring resource utilization and execution time vs running just on CPU and GPU. Analyze the impact of overhead of running both GPU and CPU on a single machine and the benefits of parallelism.
\end{enumerate}

\subsubsection{Simulation Parameters}

\begin{itemize} 
    \item \textbf{Simulation variants}: 
    \begin{itemize} 
        \item \textit{BOX}: 32, 128, 256, 512, 1024, 2056, 4096, 8192, 16384, 32768, 65536, 131072, 256000, 512000 
        \item \textit{BOX\_AND\_BALL}: 32, 128, 256, 512, 1024, 2056, 4096, 8192, 16384, 32768, 65536, 131072, 256000, 512000 
        \item \textit{ARM\_WITH\_ROPE}: 32, 128, 256, 512, 1024, 2056, 4096, 8192, 16384, 32768, 65536, 131072, 256000 
        \item \textit{HUMANOID}: 32, 128, 256, 512, 1024, 2056, 4096, 8192, 16384, 32768 
    \end{itemize} 
    \item \textbf{Simulation Steps}: 1000 
    \item \textbf{Repetitions}: 3 
\end{itemize}

\FloatBarrier 
\subsection{Results}

The following measures are tracked:

\begin{itemize}
    \item \textbf{Sequential CPU}: Running simulation across variants on CPU - also in Figure~\ref{fig:cpu_gpu_variants} .
    \item \textbf{Sequential GPU}: Running simulation across variants on GPU - also in Figure~\ref{fig:cpu_gpu_variants} .
    \item \textbf{Naive Sum}:  CPU time + GPU time given the number of variants. This curve is dominated by the slowest of standalone GPU CPU and does not benefit from parallelism. However, this metric omits the overhead of initializing and running both hardware in parallel.
    \item \textbf{Combined}: Wall clock time of running simulation across both GPU and CPU with variants allocated based on benchmark statistics from a trial run. This measure benefits from parallelism, but also accounts for the overhead of communication between CPU and GPU.
    \item \textbf{GPU variants}: Variant percentage allocated to GPU for a given simulation
\end{itemize}

\begin{figure}[hbtp]
    \centering
    \begin{minipage}{0.45\textwidth}
        \centering
        \includegraphics[width=\textwidth]{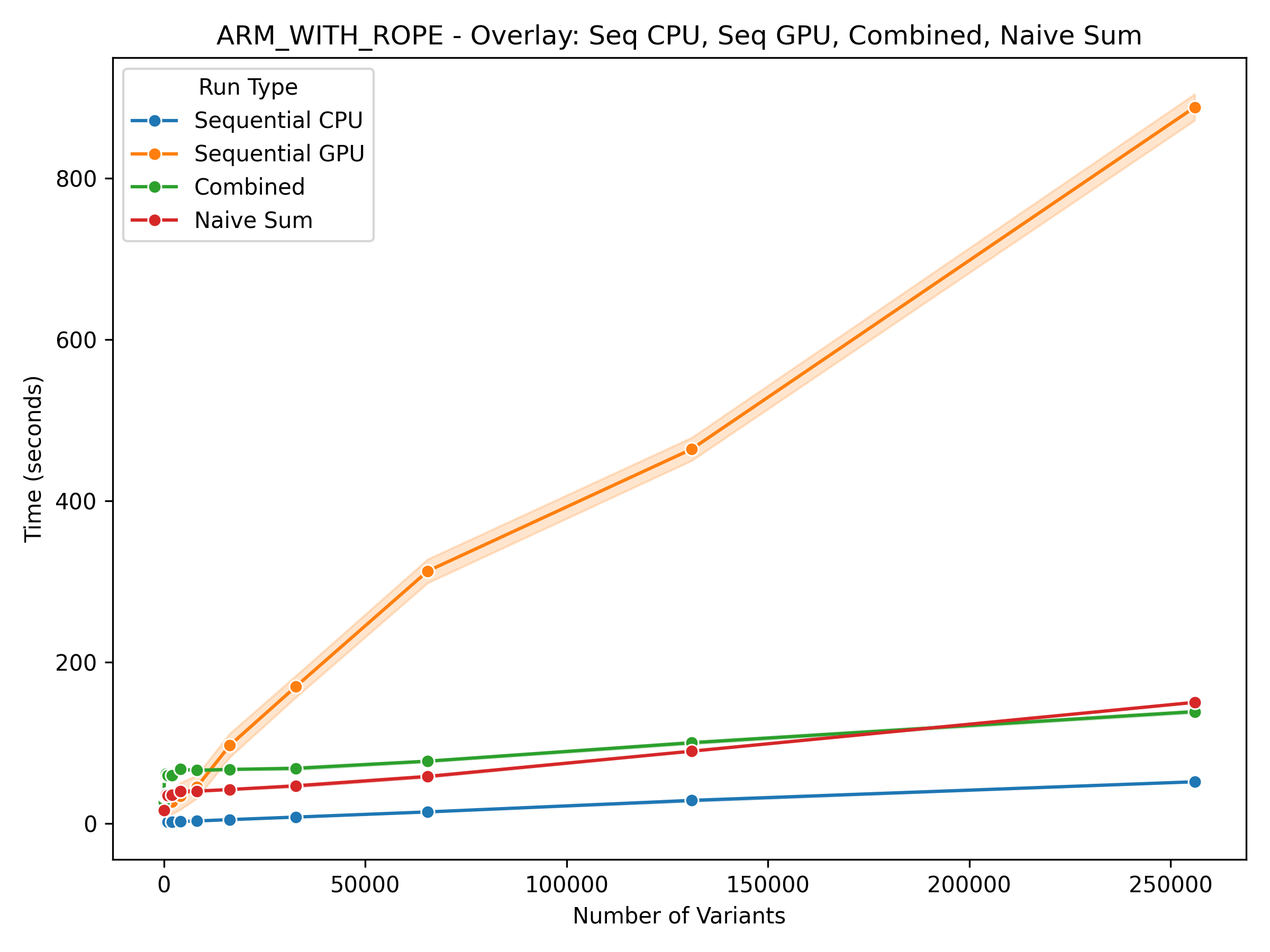}
        \label{fig:line_arm_with_rope}
    \end{minipage}
    \hfill
    \begin{minipage}{0.45\textwidth}
        \centering
        \includegraphics[width=\textwidth]{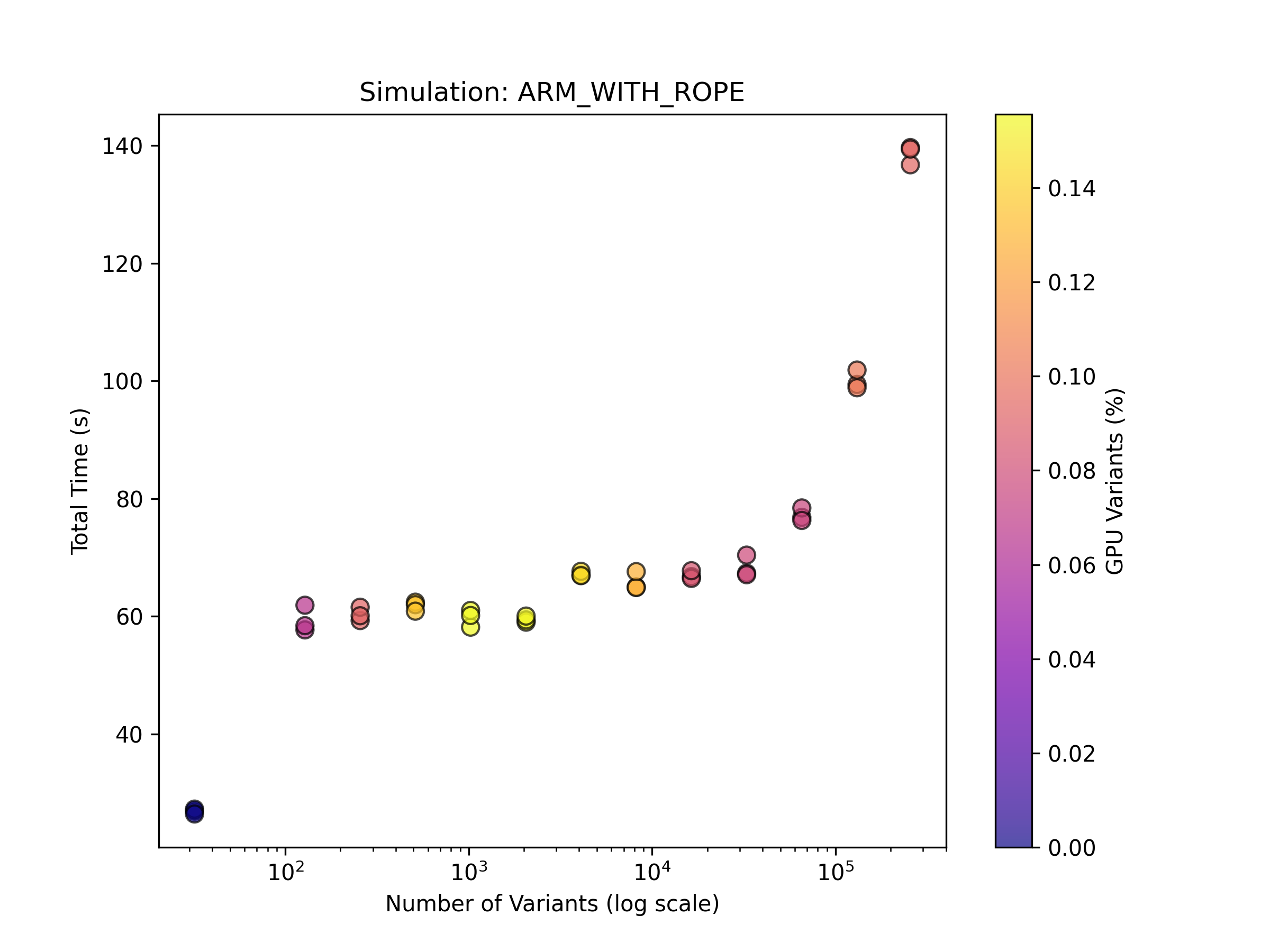}
        \label{fig:overlay_arm_with_rope}
    \end{minipage}
    \vspace{1em}
    \begin{minipage}{0.45\textwidth}
        \centering
        \includegraphics[width=\textwidth]{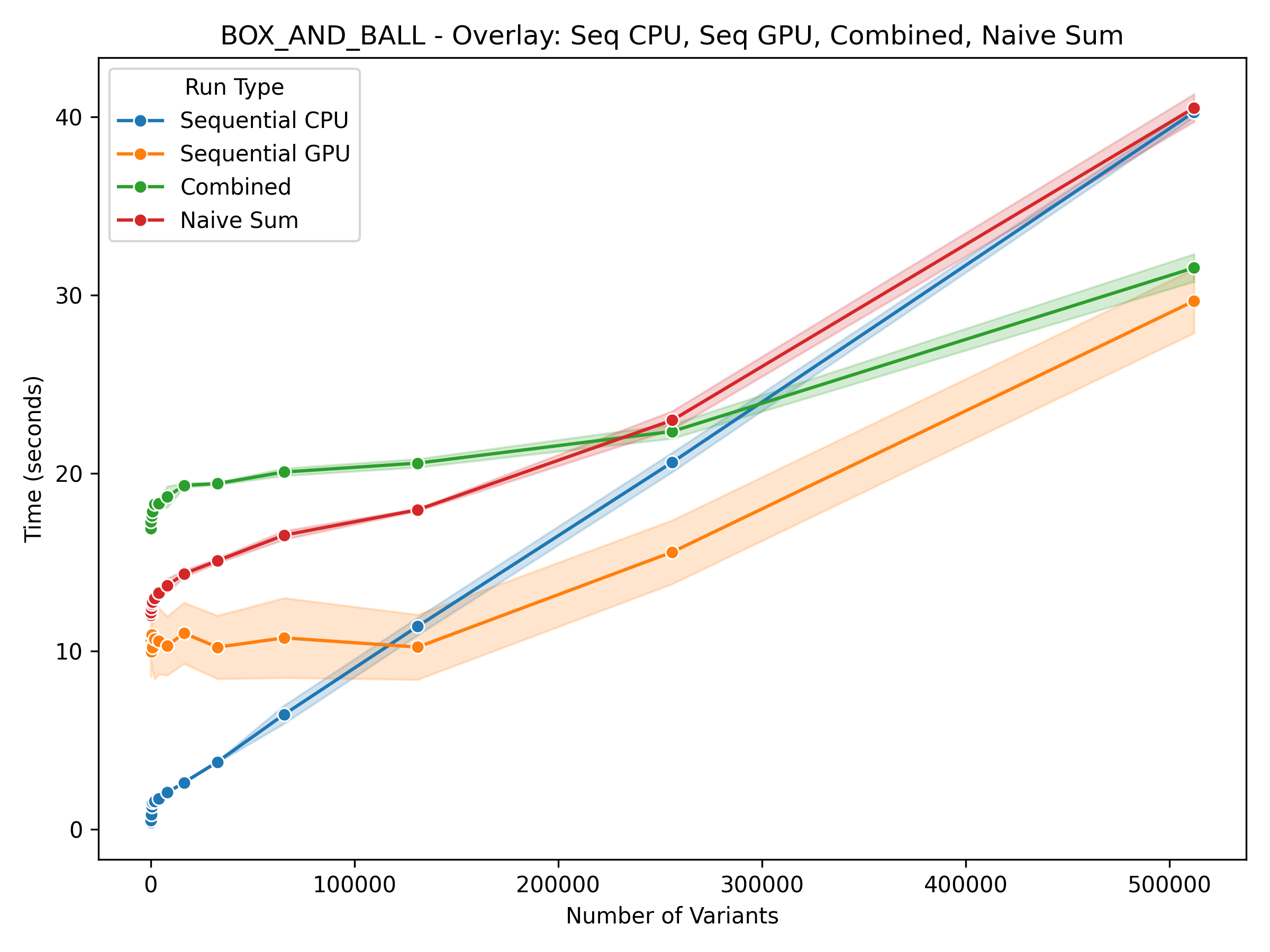}
        \label{fig:line_box_and_ball}
    \end{minipage}
    \hfill
    \begin{minipage}{0.45\textwidth}
        \centering
        \includegraphics[width=\textwidth]{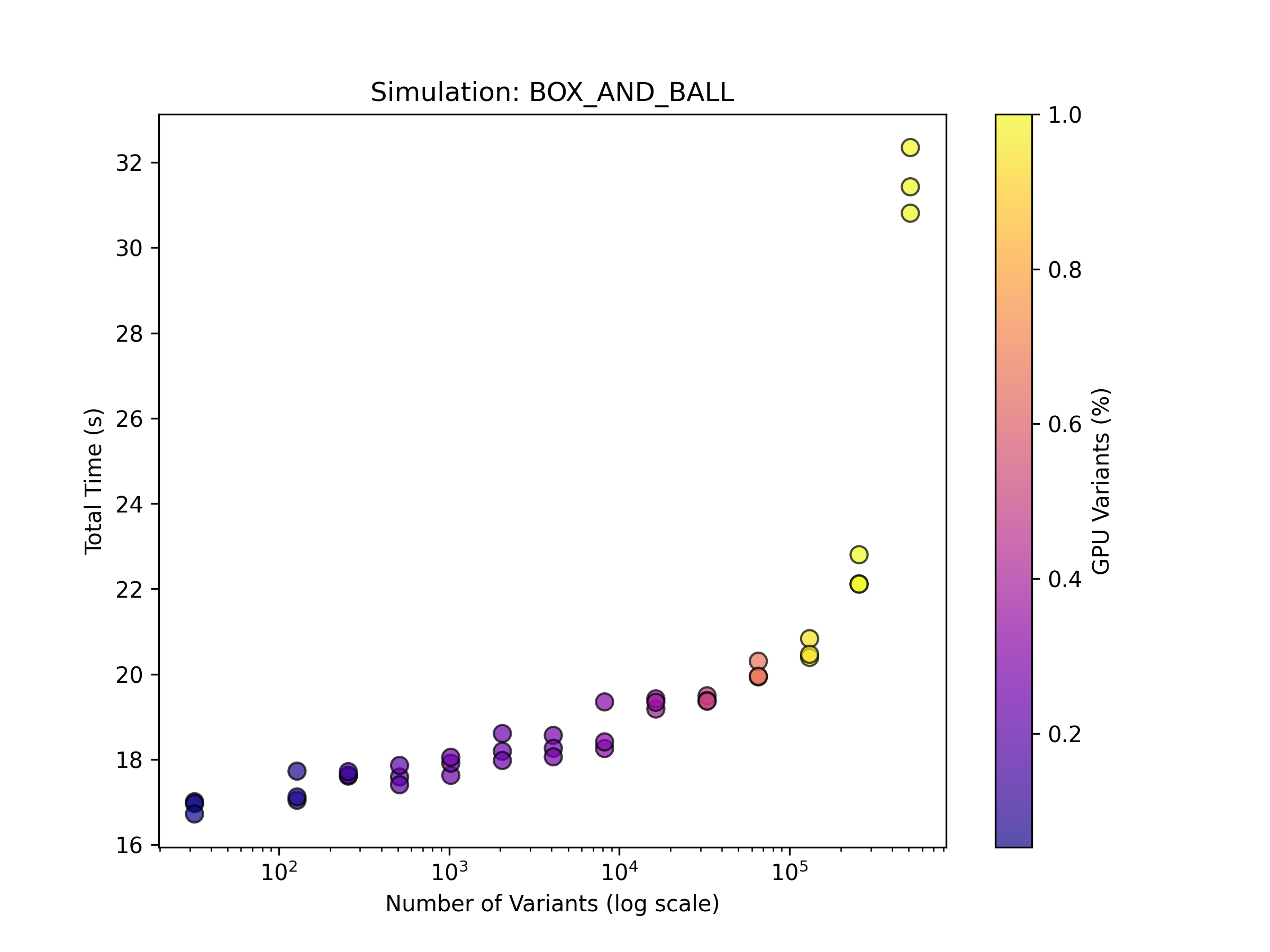}
        \label{fig:overlay_box_and_ball}
    \end{minipage}
    \vspace{1em}
    \begin{minipage}{0.45\textwidth}
        \centering
        \includegraphics[width=\textwidth]{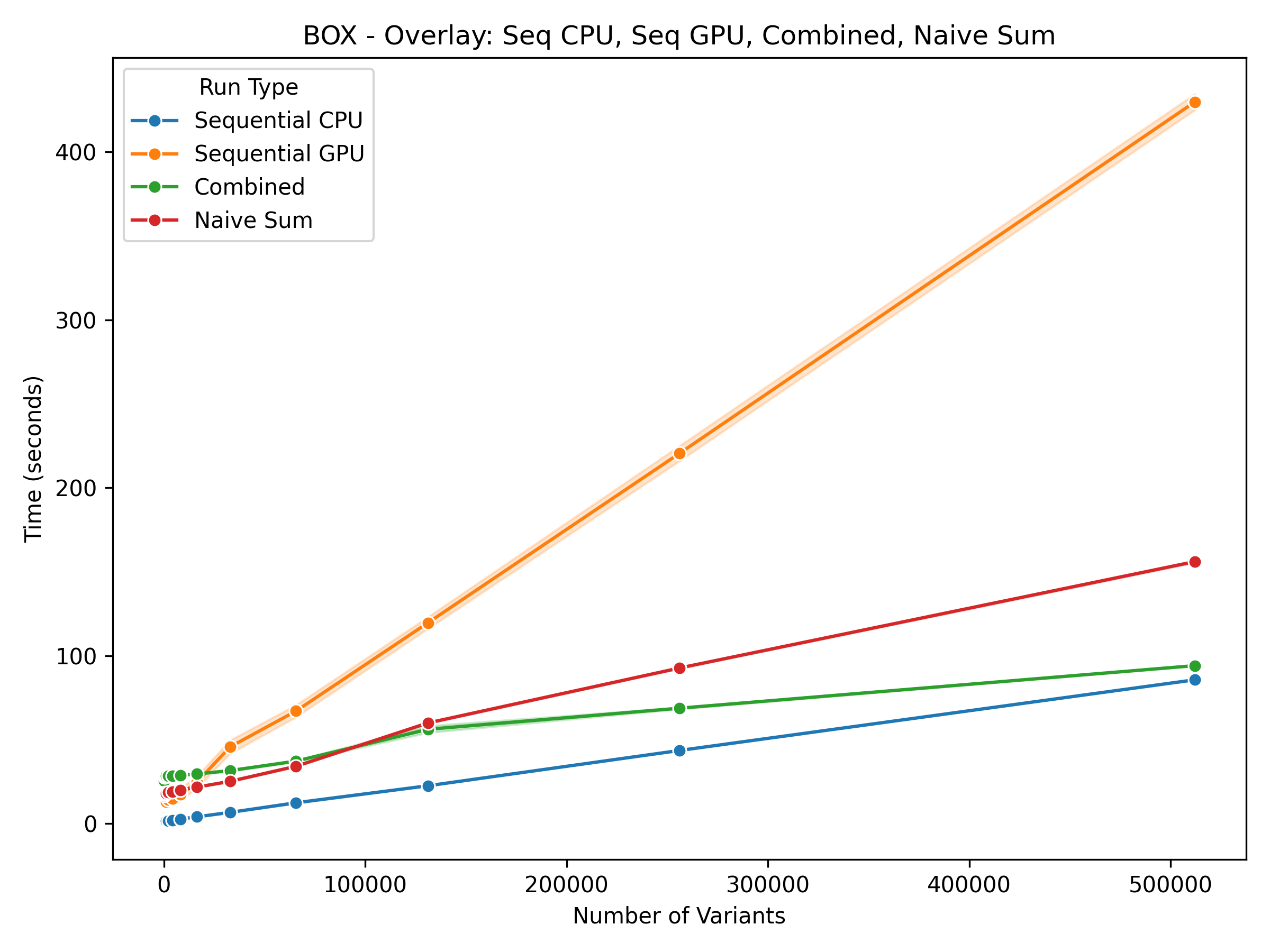}
        \label{fig:line_box}
    \end{minipage}
    \hfill
    \begin{minipage}{0.45\textwidth}
        \centering
        \includegraphics[width=\textwidth]{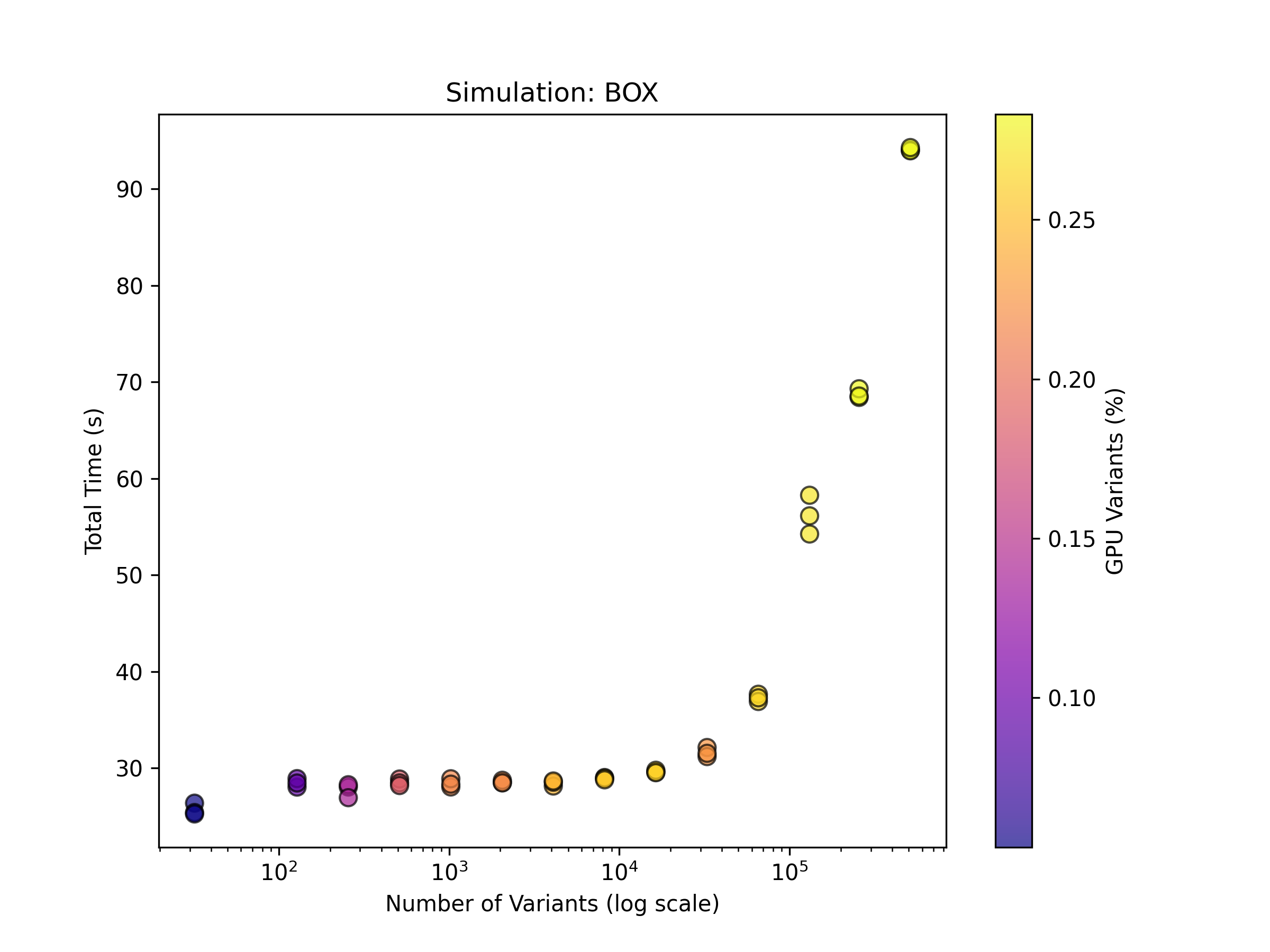}
        \label{fig:overlay_box}
    \end{minipage}
    \vspace{1em}
    \begin{minipage}{0.45\textwidth}
        \centering
        \includegraphics[width=\textwidth]{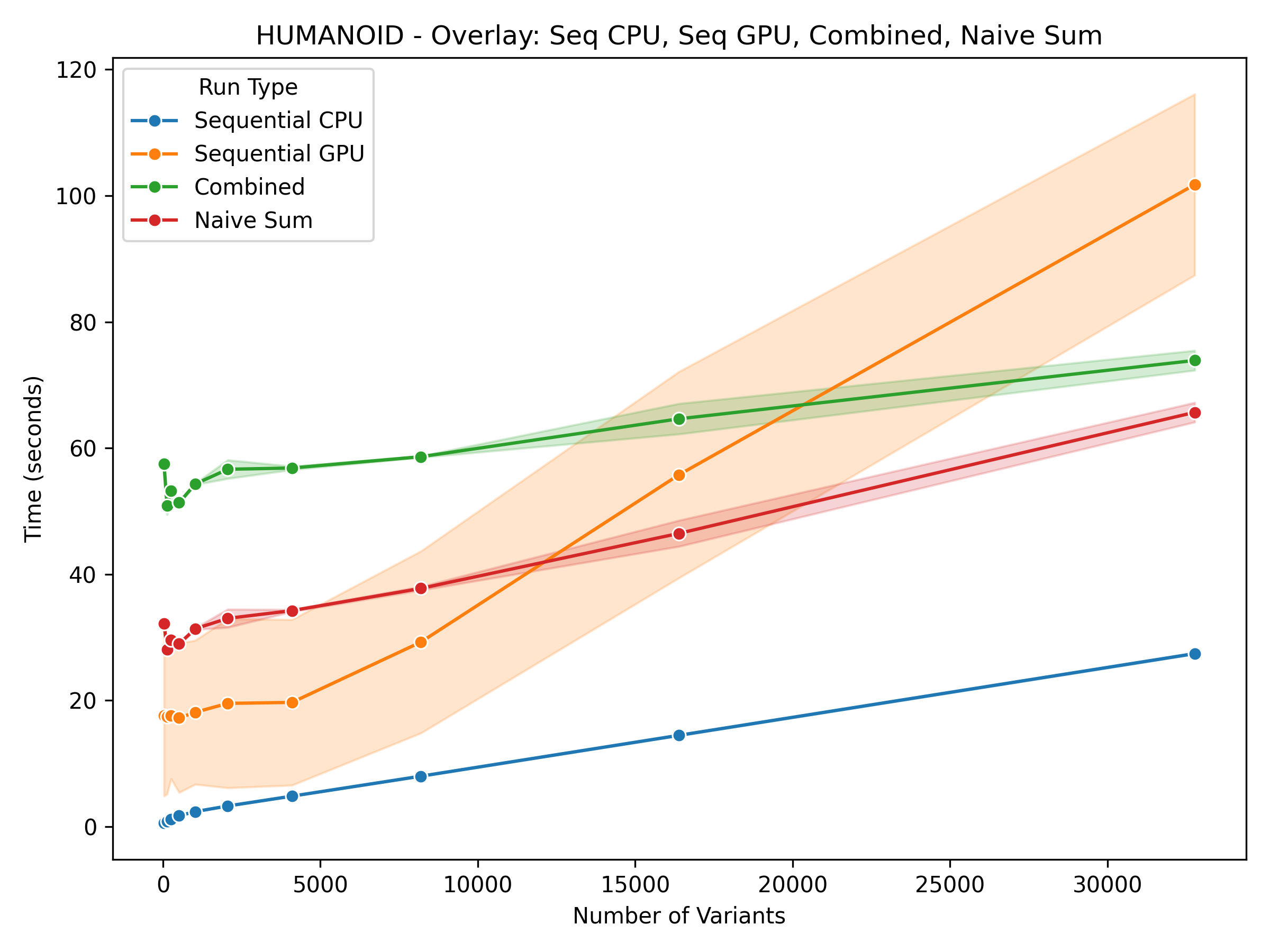}
        \label{fig:line_humanoid}
    \end{minipage}
    \hfill
    \begin{minipage}{0.45\textwidth}
        \centering
        \includegraphics[width=\textwidth]{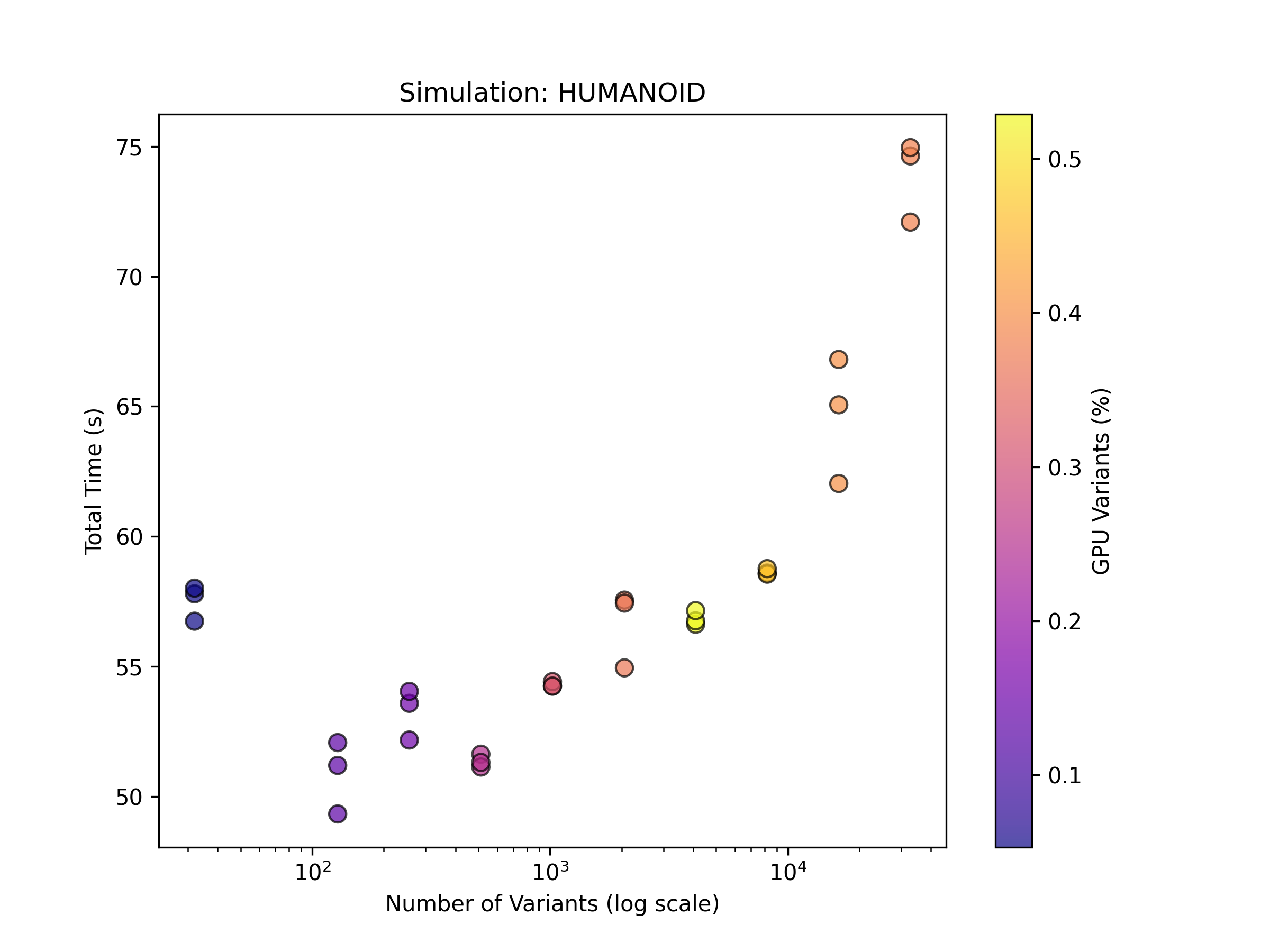}
        \label{fig:overlay_humanoid}
    \end{minipage}
    \caption{Comparison of Sequential vs Combined Runs for different models, showing both line charts and GPU percentages.}
    \label{fig:sequential_combined}
\end{figure}

\subsection{Key Observations}

\begin{itemize}
    \item \textbf{CPU Out performance}: In most scenarios, CPU only simulations were by far the fastest.
    \item \textbf{Benefits of Parallelism overshadowed by overhead cost on lower numbers of variants}: In most scenarios, the actual wall clock time delineated by the green line was higher than the naive addition of individual CPU and GPU run times. This is unexpected because the Naive sum is dominated by the slowest of the 2 hardware options. This indicates that the overhead involved completely negates the performance benefits of running the GPU and CPU in parallel.
    \item \textbf{Combined Strategy approaches the fastest hardware only run times at high variants}: In some simulations e.g. BOX, BOX\_AND\_BALL the combined strategy is approaching the dominant hardware alone. In the case of OX, BOX\_AND\_BALL, the only simulation where GPU performs better than CPU, it seems to be on track to outperform the GPU. In BOX simulation, It's approaching the CPU performance. Unfortunately, we were unable to test this simulation on further, larger numbers of variants due to the lack of sufficient memory. However, the trend lines indicate that at larger variant numbers the combined strategy can indeed surpass both CPU and GPU-only strategy in some scenarios.
\end{itemize}
\FloatBarrier 
\section{Discussion}

\subsection{Conclusion}

The results highlight the challenges of GPU-based acceleration for Mujoco simulations.
Instead, a far more promising avenue for speeding up Mujoco Simulations would be parallelizing the simulation runs across a cluster of multi-core CPUs.

\subsection{Future Work}
One possible direction for future work would be benchmarking simulations on a wider range of hardware with different configurations. For example, latest GPUs with significantly higher number of cores and machines with larger memory seem promising as the performance curves of the combined strategy seem to have a flatter slope than either GPU or CPU only variants. Still it seems naively scaling the workload across a cluster of CPUs to be the most promising solution that should scale near linearly with the number of CPUs as Evolutionary algorithms require minimal communication across the machines.
Another possible solution would involve using a different simulator that would offer better GPU utilization. 

\section*{Acknowledgments}
We would like to express our gratitude to Professor Eiben from Vrije Universiteit Amsterdam for his guidance in our research. Additionally, we thank Computational Intelligence Group for providing access to the hardware used in experiments. Lastly, we would like to express gratitude to Kaleem Ullah for helping with resolving dependency and incompatibility issues and Claire Shen for coming up with idea of benchmarking MJX outside of Revolve2 framework and the first implementation of benchmarking script.

\section*{Author Contributions}
Rustam Eynaliyev was responsible for designing and conducting the experiment, code, gathering data, implementing combined CPU+ GPU utilization scheme, collecting and visualizing the data as well as writing Sections 2-8 of this paper. Houcen Liu was responsible for writing the abstract and introduction part of the paper and also for the arrangement of the references. Both authors reviewed the manuscript and approved the final version for submission.

\end{document}